\documentclass[preprint,onecolumn,showpacs,amsmath,amssymb]{revtex4}

\usepackage{graphicx}
\usepackage{dcolumn}
\usepackage{bm}
\usepackage[squaren]{SIunits}
\usepackage{amssymb}
\usepackage{color}

\begin{document}
\title{Structure and Physical properties of New layered oxypnictides $Sr_4Sc_2O_6M_2As_2$ (M=Fe and Co) }
\author{Y. L. Xie, R. H. Liu, T. Wu, G. Wu, Y. A. Song, D. Tan, X. F. Wang, H. Chen, J. J. Ying, Y. J. Yan, Q. J. Li}
\author{X. H. Chen}
\altaffiliation{Corresponding author} \email{chenxh@ustc.edu.cn}
\affiliation{Hefei National Laboratory for Physical Science at
Microscale and Department of Physics, University of Science and
Technology of China, Hefei, Anhui 230026, China\\}
\date{\today}

\begin{abstract}
We have successfully prepared the new layered oxypnictides
$Sr_4Sc_2O_6M_2As_2$ (M=Fe and Co). They adopt the tetragonal
structure, being the same as that of $Sr_4Sc_2O_6Fe_2P_2$. The
lattice constants are a=0.4045 nm and c=0.5802 nm for M=Fe, and
a=0.4045 nm and c=1.5695 nm for M=Co, respectively. Their
transport and magnetic properties have been systematically
studied. The temperature dependence of Hall coefficient and
thermoelectric powder for $Sr_4Sc_2O_6Fe_2As_2$ compound show
complicated behavior, similar to that of iron-based parent
compounds LnOFeAs and $BaFe_2As_2$. It suggests that the
$Sr_4Sc_2O_6Fe_2As_2$ could be considered a new parent compound as
iron-based superconductors.

\end{abstract}


\maketitle

\section{Introduction}

  The recent discovery of superconductivity in iron-based arsenide with
the critical temperature ($T_C$) higher than McMillan limit of 39 K
(the theoretical maximum predicted by BCS
theory)\cite{yoichi,chen,ren,rotter} has generated great excitement
since the superconductivity is clearly unconventional similar to the
cuprate superconductors. The iron-based arsenide superconductors
have a quasi-two dimensional structure, in which Fe$_2$As$_2$ layers
are separated by different charge reservior, such as LnO (Ln = La,
Sm, Pr, Ce, Gd etc.)and Ae (Ae = Ba, Ca, Sr etc.).

To attempt new iron-based superconductor with higher critical
temperature, more and more new iron-based compounds with different
structures have been found and superconductivity was also found in
these system, such as LiFeAs, FeSe etc\cite{wang,Pitcher,Tapp,Hsu}.
However, superconductivity was not reported for the attempts to
replace old charge reservior with thicker ones, such as
Sr$_3$Sc$_2$O$_5$Fe$_2$As$_2$\cite{zhu}. Increasing the thickness of
charge reservior can make the system more two-dimensional and
enhance the spin fluctuation and density of state at Fermi surface,
consequently the superconducting transition temperature can be
pushed to higher. Recently, a similar attempt was successful in FeP
materials\cite{ogino}. Ogino et al. found that the
Sr$_4Sc_2O_6Fe_2P_2$ is superconducting below 17K, which is higher
than LaOFeP with T$_c$ $\sim$ 5K\cite{Kamihara}. This new material
has a similar structure with Sr$_4Sc_2O_6Cu_2S_2$, which has been
reported earlier\cite{Hor}. In this new FeP material, the charge
reservior is thicker than the one of LaOFeP, so that distance
between the conducting FeP layers is much increased. It indicates
that an increase in the thickness of charge reservior should enhance
the superconducting critical temperature. In the other hand, the
critical transition temperature of doped FeAs systems are much
higher than the one of FeP system with similar
structure\cite{Kamihara,yoichi}. Above findings shed light for
finding new superconductor with higher critical transition
temperature by searching new FeAs/FeP compounds with thicker charge
reservior.

In our paper, we successfully prepared New layered oxypnictides
$Sr_4Sc_2O_6M_2As_2$ (M=Fe and Co). The two materials adopt the same
crystalline structure with Sr$_4Sc_2O_6Fe_2P_2$. Their transport and
magnetic properties have been systematically studied. The
temperature dependence of Hall coefficient and thermoelectric powder
for $Sr_4Sc_2O_6Fe_2As_2$ compound show complicated behavior,
similar to that of iron-based parent compounds LnOFeAs and
$BaFe_2As_2$. It suggests that the $Sr_4Sc_2O_6Fe_2As_2$ could be
considered a new parent compound as iron-based superconductors.

\section{Experiment}

Polycrystalline samples of Sr$_4$Sc$_2$O$_6$M$_2$As$_2$ (M=Fe and
Co)were synthesized by solid state reaction method using SrAs,
Fe$_2$As, Co$_2$As, SrO and Sc$_2$O$_3$ as starting materials.
Fe$_2$As and Co$_2$As was pre-synthesized by heating the mixture
of Fe and As powder at 900$\celsius$ for 24 hours. SrAs was
prepared by the reaction of Sr lumps and As powder at
800$\celsius$ for 24 hours. The raw materials were accurately
weighed according to the stoichiometric ratio of
Sr$_4$Sc$_2$O$_6$M$_2$As$_2$ (M=Fe and Co), then thoroughly
grounded and pressed into pellets. The pellets were wrapped with
Ta foil and sealed in evacuated quartz tube. The sealed tubes were
sintered at 900$\celsius$ for 10 hours and 1200$\celsius$ for 50
hours. The sample preparation process except for annealing was
carried out in glove box in which high pure argon atmosphere is
filled. X-ray diffraction was carried out on an 18 kW MXPAHF X-ray
diffractometer with high-intensity Cu Ka radiation. Structural
refinement was performed using the analysis program GSAS. The
resistance was measured by an AC resistance bridge(LR-700, linear
Research). Magnetic susceptibility measurements were performed
with a supercondcucting quantum interference device magnetometer
(Quantum Design MPMS-7)

\section{Results and Discussions}

X-ray powder diffraction patterns are shown in Fig.1(a) for the
polycrystalline samples: Sr$_4$Sc$_2$O$_6$Fe$_2$As$_2$ and
Sr$_4$Sc$_2$O$_6$Co$_2$As$_2$, respectively. Nearly all diffraction
peaks in the patterns of both samples can be indexed by the
tetragonal with the space group P4/nmm, indicating that the samples
are almost single phase. The lattice parameters are a = 0.4045 nm
and c = 1.5802 nm for the sample Sr$_4$Sc$_2$O$_6$Fe$_2$As$_2$ and a
= 0.4045 nm and c = 1.5695 nm for the sample
Sr$_4$Sc$_2$O$_6$Co$_2$As$_2$, respectively. The crystal structure
for both of samples is shown in Fig.1(b), being similar to the
Sc-22426 compound with FeP layer.\cite{ogino}  New layered
oxypnictides $Sr_4Sc_2O_6M_2As_2$ (M=Fe and Co) consist of stacking
of antifluorite-type M$_2$As$_2$ layer and perovskite type
Sr$_4$Sc$_2$O$_6$ layer, and the distance between anti-fluorite
M$_2$As$_2$ layers is about 15.7$\AA$

Figure 2 shows temperature dependence of resistivity for the samples
Sr$_4$Sc$_2$O$_6$M$_2$As$_2$ (M=Fe and Co). As shown in Fig.2, the
resistivity of Sr$_4$Sc$_2$O$_6$Fe$_2$As$_2$ exhibits a complicated
temperature dependent behavior. A slightly metallic behavior shows
up above 200 K, and a weak semiconducting-like behavior is observed
with decreasing temperature down to about 50 K. Below 50 K, the
resistivity sharply increases with decreasing temperature. For the
sample Sr$_4$Sc$_2$O$_6$Co$_2$As$_2$, the resistivity show a
metallic behavior above about 100 K, and a weak semiconducting-like
behavior below 100 K.

The temperature dependent susceptibility for
Sr$_4$Sc$_2$O$_6$M$_2$As$_2$(M= Co, Fe) is shown in Fig.3. For
Sr$_4$Sc$_2$O$_6$Fe$_2$As$_2$, a well defined Curie-Weiss behavior
was observed in the whole temperature region from 4 K to 300 K, and
no obvious magnetic transition was observed down to 4K, which is
different from iron-based parent compounds of 1111-type and
122-type\cite{yoichi, rotter2}, but similar with
Sr$_3$Sc$_2$O$_5$Fe$_2$As$_2$\cite{zhu}. It seems to be that
increasing thickness of charge reservior is harmful for the
formation of spin density wave transition. This maybe ascribed to
strong magnetic fluctuation in these more two-dimensional
structures. For Sr$_4$Sc$_2$O$_6$Co$_2$As$_2$, a ferromagnetic-like
behavior was observed, but the Curie temperature maybe higher than
measurable temperature range. It is similar with LaOCoAs and
LaOCoP\cite{Yanagi,Sefat}, in which an itinerant ferromagnetism was
also observed. It is possible that the Curie temperature for
Sr$_4$Sc$_2$O$_6$Co$_2$As$_2$ is much higher than LaOCoAs (T$_c$
$\sim$ 66K).

In order to identify carrier-type for Sr$_4$Sc$_2$O$_6$Fe$_2$As$_2$,
its Hall effect and the thermoelectric power (TEP) are
systematically measured. Figure 4a shows the temperature dependence
of thermoelectric power. TEP of the sample
Sr$_4$Sc$_2$O$_6$Fe$_2$As$_2$ is negative at room temperature, and
decreases with decreasing temperature, and show an anomaly peak at
about 180 K. Such anomalous peaks in TEP coincides with the
occurrence of the crossover from metallic to semiconducting behavior
in resistivity as shown in Fig.2. With further decreasing
temperature below 180 K, the TEP increases  and the change of TEP
sign happens at about 95 K. Such TEP behavior is very similar to
that of $MFe_2As_2$ (M=Ba, Ca) and LnOFeAs.\cite{wu1,wu2} Figure 4b
shows the temperature dependence of the Hall coefficient R$_H$. One
can see that R$_H$ remains negative in the whole temperature range
from 5 K to 300 K. The absolute value of R$_H$ increases rapidly
below about 50 K at which the resistivity starts to increase sharply
with decreasing temperature. Such behavior is similar to that
observed in SmOFeAs in which SDW ordering leads to rapid increase of
Hall coefficient.\cite{liu}. Negative Hall coefficient and TEP
indicate that electron-type charge carriers dominate the conduction
in the sample. The different sign of Hall coefficient and TEP
implies that the system is multiband, being similar to the case of
LnOFeAs(1111) and BaFe$_2$As$_2$(122) parent compounds.

\section{Conclusions}
New layered oxypnictides $Sr_4Sc_2O_6M_2As_2$ (M=Fe and Co) are
successfully discovered. They adopt the tetragonal structure, being
the same as that of $Sr_4Sc_2O_6Fe_2P_2$.  Their transport and
magnetic properties have been systematically studied. The
temperature dependence of Hall coefficient and thermoelectric powder
for $Sr_4Sc_2O_6Fe_2As_2$ compound show complicated behavior,
similar to that of iron-based parent compounds LnOFeAs and
$BaFe_2As_2$. Our studies suggest that Sr$_4$Sc$_2$Fe$_2$As$_2$O$_6$
maybe a new parent compound for iron-based superconductor.

{\bf Note:} When we are preparing this manuscript, the competing
work is reported by G. F. Chen et al., arXive:0903.5273 and H. Ogino
et al., arXive:0903.5124.

\newpage

\begin{figure*}[t]
\includegraphics[width=15cm]{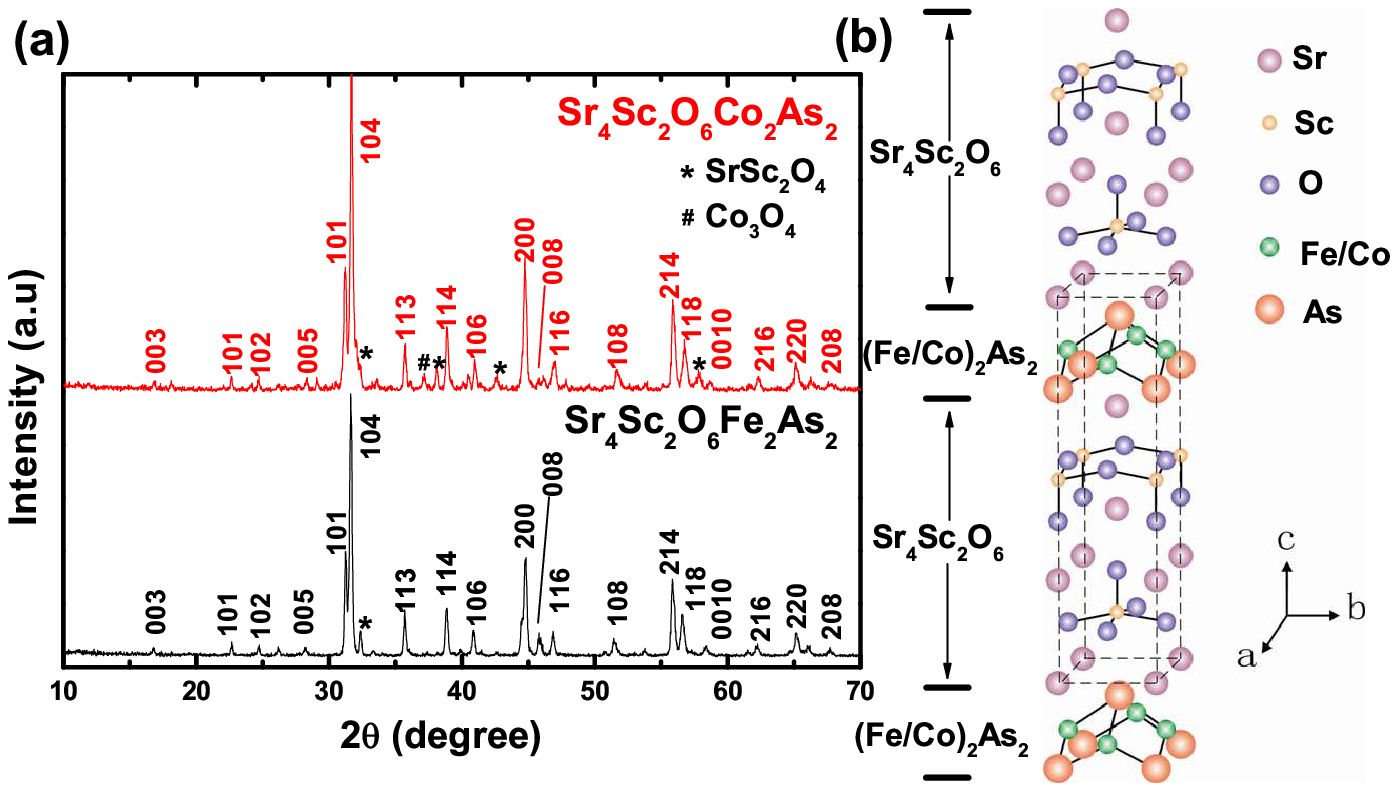}
\caption{(Color online) (a). X-ray powder diffraction patterns for
Sr$_4Sc_2O_6Co_2As_2$ and Sr$_4Sc_2O_6Fe_2As_2$; (b). Crystal
structure for Sc-22426.}
\end{figure*}

\begin{figure*}[t]
\includegraphics[width=15cm]{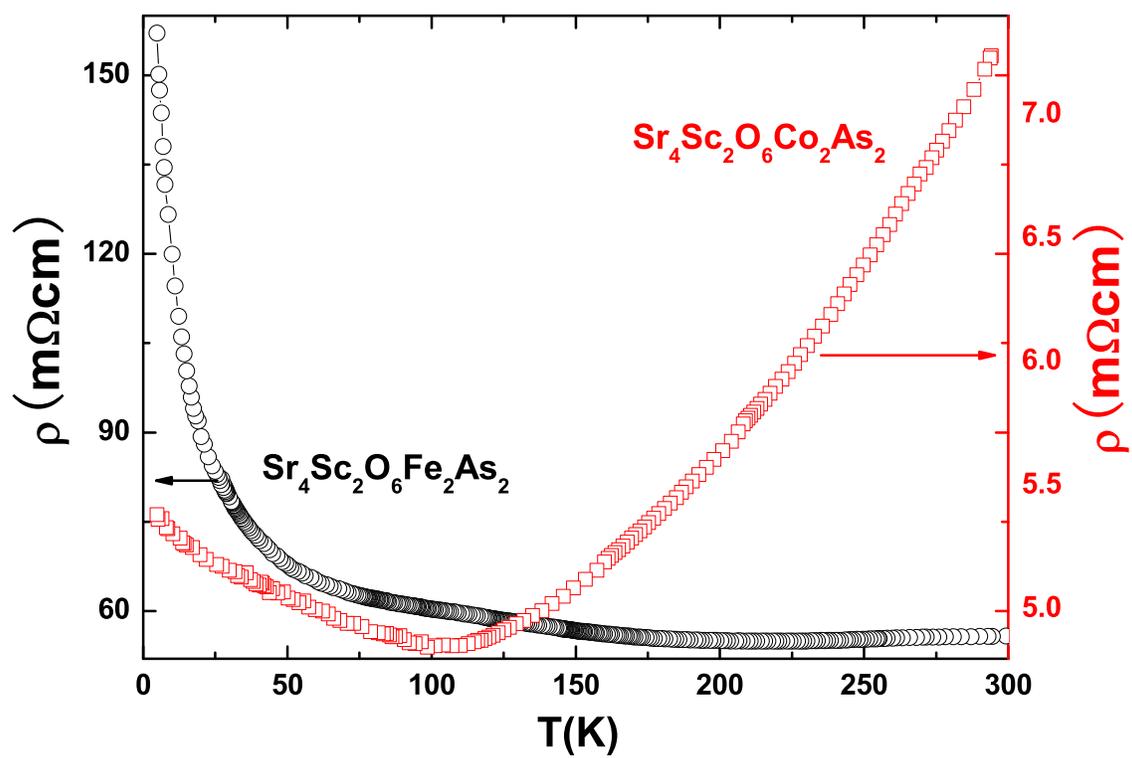}
\caption{Temperature dependence of resistivity for the samples
Sr$_4$Sc$_2$O$_6$M$_2$As$_2$(M=Fe and Co)}
\end{figure*}

\begin{figure*}[t]
\includegraphics[width=15cm]{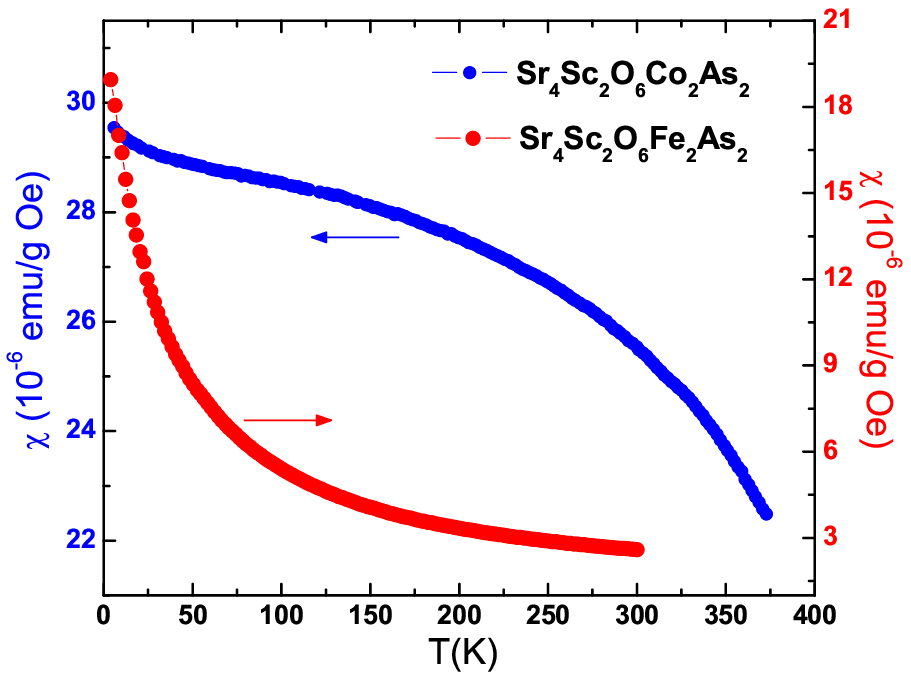}
\caption{Temperature dependence of susceptibility for the samples
Sr$_4$Sc$_2$O$_6$M$_2$As$_2$(M=Fe and Co)}
\end{figure*}

\begin{figure*}[t]
\includegraphics[width=15cm]{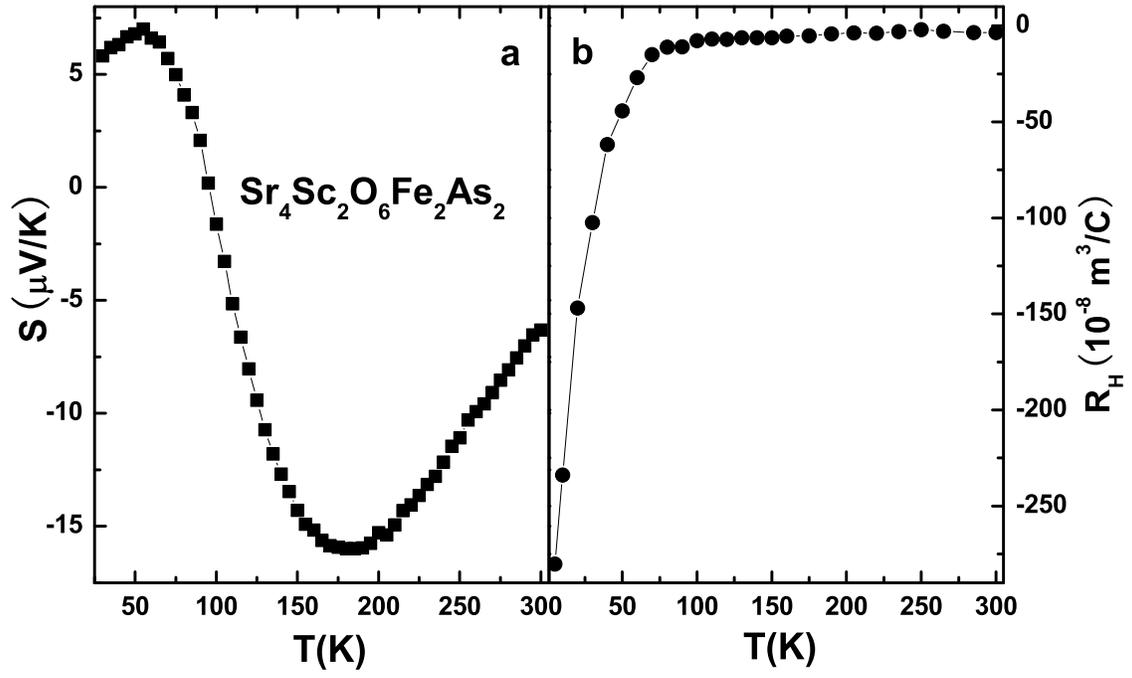}
\caption{Temperature dependence of thermoelectric power and Hall
coefficient R$_H$ for the samples }
\end{figure*}

\end{document}